\documentclass[twocolumn,letterpaper]{IEEEAerospaceCLS}

\usepackage{amsmath,amssymb}
\usepackage{booktabs}
\usepackage{graphicx}
\setkeys{Gin}{keepaspectratio}
\usepackage{multirow}
\usepackage{url}
\usepackage[hidelinks]{hyperref}
\newcommand{\godsat}{\textsc{GoDSAT}}

\title{Constrained Diffusion for Data-Scarce Orbital Monte Carlo in Constellation Tasking}

\author{%
\begin{tabular*}{.96\textwidth}{@{\extracolsep{\fill}}ccc@{}}
\begin{tabular}{c}
\textbf{Omar Ramadan}\\
Whiting School of Engineering\\
Johns Hopkins University\\
3400 N. Charles St.\\
Baltimore, MD 21218, USA\\
oramada1@jhu.edu
\end{tabular}
&
\begin{tabular}{c}
\textbf{Sam Siavoshian}\\
Whiting School of Engineering\\
Johns Hopkins University\\
3400 N. Charles St.\\
Baltimore, MD 21218, USA\\
samsiavoshian2009@gmail.com
\end{tabular}
&
\begin{tabular}{c}
\textbf{Amir Kashif Saeed}\\
Whiting School of Engineering\\
Johns Hopkins University\\
3400 N. Charles St.\\
Baltimore, MD 21218, USA\\
asaeed7@jhu.edu
\end{tabular}
\\[3em]
\begin{tabular}{c}
\textbf{Benjamin A. Johnson}\\
Whiting School of Engineering\\
Johns Hopkins University\\
3400 N. Charles St.\\
Baltimore, MD 21218, USA\\
benjamin.a.johnson@jhu.edu
\end{tabular}
&
\begin{tabular}{c}
\textbf{Amin M. E.-A. Diab}\\
Whiting School of Engineering\\
Johns Hopkins University\\
3400 N. Charles St.\\
Baltimore, MD 21218, USA\\
adiab3@jhu.edu
\end{tabular}
&
\begin{tabular}{c}
\textbf{Benjamin M. Rodriguez}\\
Whiting School of Engineering\\
Johns Hopkins University\\
3400 N. Charles St.\\
Baltimore, MD 21218, USA\\
brodrig5@jhu.edu
\end{tabular}
\end{tabular*}
}

\begin{document}
\maketitle
\thispagestyle{plain}
\pagestyle{plain}

\noindent{\footnotesize\textit{This work has been submitted to the IEEE for possible publication. Copyright may be transferred without notice, after which this version may no longer be accessible.}}\par\medskip

\begin{abstract}
Constellation Monte Carlo studies depend on both a tasking policy and the orbital trajectories used to evaluate that policy. When reference trajectories are scarce, seed replay increases campaign count without expanding handoff geometry, while independent orbital-element jitter can produce dynamically inconsistent or Earth-intersecting orbits. We investigate whether constrained diffusion can expand a scarce orbital population for constellation tasking while preserving physical executability. A force-conditioned denoising diffusion model learns a 13-dimensional orbital prior in which semimajor axis is recovered analytically from perigee altitude and eccentricity. Each generated sample is propagated in Basilisk under one of five force-model tiers, separating learned population variability from simulated temporal evolution. We compare diffusion with jittered bootstrap, per-tier Gaussian mixtures, and a conditional variational autoencoder using an 800-trajectory corpus. Evaluation includes held-out and shifted-blind splits, a Lambert interior check, physics-residual estimation diagnostics, and 4,000 paired GoDSAT-compatible campaigns. On held-out trajectories, the largest diffusion model achieves maximum mean discrepancy of 0.0171 ± 0.0239, with similar held-out MMD to bootstrap (0.0170) and the mixture (0.0198), while generating samples farther from individual training priors (median nearest-training distance 2.7 versus 0.10 standardized units). All six generators remain readily distinguishable from the shifted-blind target population in every seed–tier block (classifier AUC 0.991–0.998). Endpoint-conditioned samples meet Lambert boundary conditions but show larger interior position error than the matched Lambert reference (29.2 versus 5.7 km mean RMSE). Residual diffusion improves selected sparse forecasts and catalog-mode recall but does not outperform classical estimators on custody ranking. In a common 16-satellite configuration, diffusion has similar mean custody (0.463) to seed replay (0.470) and bootstrap (0.461), while the force-tier mixture shifts custody by approximately 5.5 percentage points. Constrained diffusion is therefore useful for local, in-support Monte Carlo augmentation but cannot replace orbital dynamics or serve as an operational posterior. Trajectory-population construction should be reported as a consequential source of uncertainty in constellation analysis.
\end{abstract}


\tableofcontents

\section{Introduction}

\PARstart{S}{atellite} constellation sensor-tasking systems are commonly evaluated through Monte Carlo simulation, yet the credibility of those campaigns depends on the trajectory population supplied to the simulator. Custody transfer, tracking continuity, and handoff timing are not intrinsic scores of a tasking policy. They are properties of an interaction between that policy and a set of orbits, sensor geometries, and observation opportunities. If the supplied set is too small, too repetitive, or geometrically inconsistent, a campaign can report a stable custody number that is an artifact of the seed catalog rather than a property of the logic under test.

When reference trajectories are scarce, two default remedies fail in opposite ways. Deterministic replay preserves physical consistency but does not expand handoff geometry, pass duration, or sensing conditions. Independently jittering classical orbital elements appears to restore diversity, yet element-wise noise does not respect the joint structure of the seed set and can produce Earth-intersecting perigees or force-model settings the campaign never intended to represent. The practical need is therefore narrower than a request to generate more orbits. It is to expand a scarce, physically admissible population inside a declared simulation support, and to measure whether that expansion changes constellation-level conclusions.

This problem is directly relevant to \godsat{}, a reinforcement-learning framework for constellation sensor tasking \cite{godsat2026}. GoDSAT evaluations of tracking continuity and custody transfer are only as informative as the trajectories presented to the campaign. A generator that copies training rows inflates sample size without expanding tested handoff conditions. A generator that leaves the observed support, or that emits dynamically inconsistent state sequences, contaminates the campaign with cases the declared force model never intended to represent.

Diffusion models can represent non-Gaussian conditional distributions through iterative denoising \cite{sohldickstein2015diffusion,ho2020ddpm,song2021score} and have been applied to trajectory modeling, forecasting, and inverse problems \cite{janner2022diffuser,shen2023timediff,chung2023dps}. That literature establishes statistical flexibility. It does not establish that a learned sampler is an acceptable source of Monte Carlo orbits. Astrodynamics requires geometric admissibility and dynamic executability under a named force model, and it requires learned corrections to be compared with classical propagation, filtering, smoothing, orbit determination, and boundary-value solutions rather than treated as substitutes for them \cite{vallado2013astrodynamics,kenneally2020basilisk}. A trajectory can be statistically close to a training cloud and still be useless for constellation evaluation if it is not propagateable, if it leaves the declared support, or if it changes downstream custody only by making the campaign easier.

This paper asks a single organizing question:
\emph{Can constrained diffusion expand a data-scarce orbital trajectory population for constellation Monte Carlo analysis, and where does that augmentation cease to be reliable relative to classical astrodynamics?}
The primary generator learns a force-conditioned distribution over a 13-dimensional orbital prior and delegates temporal evolution to Basilisk. Semimajor axis is recovered from perigee altitude so that valid marginals cannot combine into an Earth-intersecting orbit. Observation-conditioned residual models and a matched Lambert test are used only as boundary diagnostics, not as competing claims that the network has learned orbital dynamics.

A central reason to evaluate diffusion, despite the low dimensionality of the prior, is that population augmentation is not the same task as matching held-out statistics. A per-tier Gaussian mixture or a jittered bootstrap can be a strong density estimator in a 13-dimensional, comparatively smooth space. Bootstrap methods can achieve high local fidelity largely through near-replication of observed cases. A learned generator may produce configurations farther from individual seeds without immediately leaving the observed support. Whether that additional diversity is admissible and decision-relevant must be measured rather than assumed. Downstream custody is therefore treated as a property of the induced population, not as a score to be maximized.

The paper contributes:
\begin{itemize}
    \item a force-conditioned orbital-prior representation that preserves angular topology and non-Earth-intersecting perigee geometry before simulator propagation;
    \item a staged validation against low-dimensional statistical generators and classical astrodynamics that separates in-support fidelity from prescribed support shift;
    \item a GoDSAT-compatible Monte Carlo study showing that trajectory-population construction can change estimated custody under a fixed tasking configuration.
\end{itemize}

The remainder of the paper defines the constraint contract and experimental protocol, reports interpolation, support shift, classical boundary tests, and the constellation campaign, and then interprets those results as a separation among physical executability, distributional fidelity, and decision quality.

\section{Related Work}

Constellation sensor tasking and catalog custody are usually studied as allocation problems: given a set of objects and a set of sensors, choose pointing, dwell, and handoff actions that keep tracks alive. Reinforcement learning and search-based schedulers have been applied to space-based and ground-based tasking, including narrow-field sensors, multi-sensor networks, and constellation-scale catalog maintenance \cite{saeed2023,roberts2022sbstask,fedeler2023mcts,georgantopoulos2022drltask}. Custody, covariance, and unique-object count are the typical figures of merit. Those papers treat the trajectory population as an input. They do not ask whether the Monte Carlo orbits used to score the policy are themselves a source of uncertainty.

\godsat{} is the evaluation setting used here \cite{godsat2026}. It is an RL framework for constellation sensor tasking in which ownership, slew, ClaimMap reservation, and custody transfer are exercised across a constellation of sensors and objects. Earlier group work studied RL tasking of LEO hyperspectral sensors and multi-agent, multi-target satellite tasking \cite{saeed2023,saeed2024multiagent}. A related multi-sensor custody architecture, GoDSAT-AT, applies the same family of ideas to terminal-airspace aircraft tracking; a separate study addresses reproducible analytics for that aircraft setting \cite{saeed2026,ramadan2026dataops}. These papers provide tasking and custody context. The present paper does not propose a new tasking policy or claim to compare those systems. It asks how a scarce orbital population should be expanded before a policy is scored.

Operational Monte Carlo practice typically perturbs two-line-element or reconstructed-state catalogs, draws from estimated covariances, or bootstraps a small set of reference arcs \cite{vallado2013astrodynamics}. Those procedures are simple and often sufficient when the catalog is dense. They become brittle when only a handful of reference trajectories exist: replay under-covers handoff geometry, and element-wise jitter can break coupled geometry or produce Earth-intersecting perigees. High-fidelity simulators such as Basilisk separate initial conditions from force models and integration, which makes them suitable as rollout engines once a prior has been sampled \cite{kenneally2020basilisk}. Classical boundary-value and estimation methods (Lambert transfers, EKF/UKF/RTS smoothing, and batch orbit determination) remain the physical references against which any learned correction must be judged \cite{vallado2013astrodynamics}.

Generative models have been used to sample spacecraft trajectories in other regimes. Score-based diffusion has been applied to ballistic Earth--Mars transfers \cite{presser2024ballistic} and to compositional generation of powered-descent trajectories \cite{briden2025diffusion6dof}. Variational autoencoders have been used to propose periodic orbits in the circular restricted three-body problem, generally followed by a shooting or continuation refinement so that the decoded curve satisfies the dynamics \cite{gil2024cr3bpvae}. Sequence diffusion has been studied for planning, forecasting, and inverse problems outside astrodynamics \cite{janner2022diffuser,shen2023timediff,chung2023dps,rasul2021timegrad}. Diffusion has also been used to approximate particle-smoothing trajectories \cite{tamir2024smoothing}. Those lines of work generate or refine \emph{trajectories}. The generator in this paper generates a constrained orbital \emph{prior} and delegates temporal evolution to a named force-model simulator. The distinction matters: statistical fit of a state sequence is not the same test as executability under Basilisk plus usefulness for constellation Monte Carlo.

Because the prior used here is only 13-dimensional and comparatively smooth, simple statistical generators are first-class baselines rather than straw men. Per-tier Gaussian mixtures and jittered bootstrap resampling are the relevant operational competitors. Distributional tests based on maximum mean discrepancy and classifier two-sample tests are used to measure whether a learned sampler is doing more than replicating training rows \cite{gretton2012mmd,lopezpaz2017c2st}. The gap addressed by this paper is therefore specific: in-support augmentation of a scarce, geometrically constrained orbital population for constellation campaign analysis, measured against those simple generators and against downstream custody, not a claim that diffusion is the preferred density model for orbital elements.

\section{Methodology}

\subsection{Problem Formulation and Monte Carlo Use Case}

Let a sparse reference set be
\begin{equation}
\mathcal{D}_{\mathrm{seed}}=\{\tau_i\}_{i=1}^{N},
\end{equation}
and let a generator produce $M\gg N$ synthetic trajectories within a declared simulation support. For the present study, an augmentation method is useful only if its samples are
(i) physically executable under the declared simulator,
(ii) statistically faithful to relevant held-out cases within support, and
(iii) informative for downstream sensitivity analysis.
These criteria concern local augmentation quality and do not imply recovery of an unknown operational population.

The observation-conditioned experiments use controlled custody surrogates to isolate estimation, uncertainty representation, and ranking behavior. Separately, the final experiment executes four frozen trajectory populations through a common GoDSAT-compatible campaign: deterministic seed replay, bootstrap resampling, force-tier Gaussian-mixture sampling followed by Basilisk repropagation, and constrained diffusion. Each population contains 1,000 trajectories. Campaign index is paired across methods through a common GoDSAT stochastic seed so that downstream contrasts isolate trajectory-population construction under shared campaign randomness.

\subsection{Trajectory Constraint Contract}

Generated trajectories are required to satisfy four distinct classes of constraints.

\emph{Geometric constraints} enforce valid orbital parameterization, angular topology, and non-Earth-intersecting perigee geometry.

\emph{Support constraints} restrict calibrated in-domain augmentation to the declared force-tier training bounds. Samples outside those bounds are treated as shifted or stress-test cases rather than calibrated population draws.

\emph{Dynamics constraints} are enforced by propagating every decoded prior through the corresponding Basilisk force model rather than independently generating position, velocity, and acceleration channels.

Finally, \emph{observation and decision constraints} describe what the sensing and custody pipeline can observe or accept. These are evaluated separately because a trajectory can be physically executable while remaining difficult for a constellation to track or maintain in custody.

Simulator execution therefore establishes consistency with the declared rollout model, not correctness of the simulator, agreement with an operational population, or validity under regime shift. Physical admissibility, distributional fidelity, extrapolation behavior, and downstream decision utility are measured separately throughout the study.

\subsection{Classical Astrodynamics and Lambert Reference}

The observation-conditioned benchmarks include two-body propagation, central-gravity extended Kalman filtering with Rauch--Tung--Striebel (EKF--RTS) smoothing, unscented Kalman filtering (UKF), regularized nonlinear batch orbit determination, and multistart estimation \cite{vallado2013astrodynamics}. These references test whether learned residuals add value after deterministic dynamics or classical estimation have already removed the dominant orbital structure.

Lambert's problem addresses a different task from unconditional population generation: given two positions and a time of flight, it returns Keplerian transfer arcs satisfying those boundary conditions. It is therefore not treated as a competing baseline for unconditional prior generation. Instead, it is used as a targeted consistency test for endpoint-conditioned trajectories.

For each of the 800 fixed reference cases, a zero-revolution Izzo Lambert solution \cite{izzo2015lambert} is constructed from the initial position, terminal position, and time of flight. When multiple valid branches exist, the branch with minimum initial-velocity distance to the reference trajectory is selected. This reference-informed rule is intentionally favorable to the controlled comparison and is not an operational branch-selection procedure.

The frozen two-body-residual temporal U-Net receives only the endpoint positions, generates one 48-step DDIM residual trajectory, and then uses a deterministic affine bridge to impose exact endpoint positions. Velocities are recomputed from the corrected positions. Both the reference trajectory and the endpoint-conditioned diffusion trajectory are compared with the same matched Lambert arc using full-trajectory position and velocity RMSE. Because several reference tiers contain perturbations beyond two-body motion, the resulting errors measure consistency with the matched Lambert solution rather than universal physical accuracy. The experiment is therefore a boundary-of-validity test: exact endpoint satisfaction is necessary for the constructed conditioning task, but it is not sufficient evidence of an interior Lambert-consistent trajectory.

\subsection{Basilisk Benchmark}

The benchmark contains 800 trajectories and 51,200 Earth-centered, Earth-fixed states generated with Basilisk 2.10.2. Splits contain 512 training, 96 validation, 96 held-out test, and 96 shifted-blind trajectories. Each trajectory contains 64 states.

Five balanced force tiers are used:
Earth point mass;
$J_2$;
GGM03S spherical harmonics through degree 8;
degree-8 gravity with exponential-atmosphere drag; and
degree-8 gravity and drag with Sun and Moon point masses evaluated from SPICE kernels.

\begin{figure*}[t]
    \centering
    \includegraphics[width=0.96\textwidth]{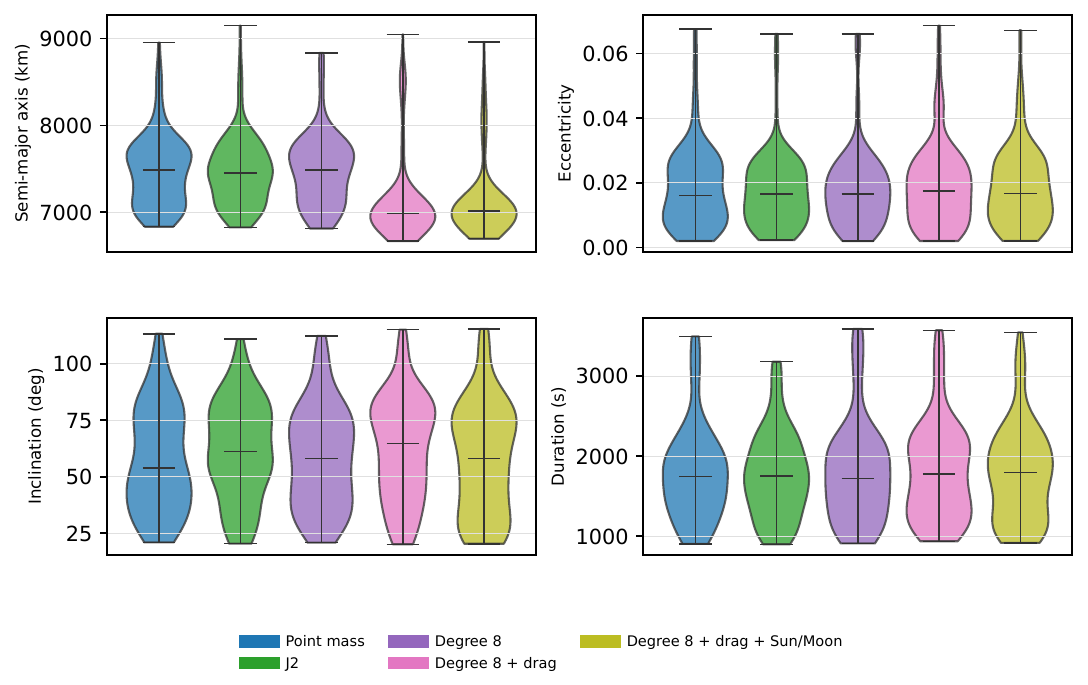}
    \caption{Coverage of the 800-trajectory Basilisk benchmark. The blind split changes orbital and duration support; orientation angles and spacecraft properties retain their declared ranges.}
    \label{fig:coverage}
\end{figure*}

The held-out test split uses the training orbital ranges. The blind split changes perigee altitude to $720$--$2170$~km, eccentricity to $0.015$--$0.070$, inclination to $72^\circ$--$116^\circ$, and duration to $2100$--$3600$~s. Figure~\ref{fig:coverage} summarizes the resulting support.

\subsection{Constrained Orbital-Prior Representation}

Each simulator input is mapped to
\begin{align}
\mathbf{z}=[&
h_p,e,i,T,m,A,C_D,\sin\Omega,\cos\Omega,\nonumber\\
&\sin\omega,\cos\omega,\sin f,\cos f],
\end{align}
where $h_p$ is perigee altitude, $e$ eccentricity, $i$ inclination, $T$ duration, $m$ mass, $A$ drag area, and $C_D$ drag coefficient. Angular sine/cosine pairs remove the discontinuity at $360^\circ$. Channels are standardized on training data and decoded within per-tier training bounds. Semimajor axis is derived analytically as
\begin{equation}
a=\frac{R_E+h_p}{1-e},
\end{equation}
which prevents individually valid marginals from combining into an Earth-intersecting perigee.

Epoch is supplied as evaluation context and excluded from learned-distribution metrics to avoid leakage from the campaign's monotonically increasing trajectory index.

The force tier
\begin{equation}
c\in\{1,\ldots,5\}
\end{equation}
conditions every neural generator. Each decoded parameter sample is propagated by the corresponding Basilisk configuration. Simulator success therefore establishes executable, constraint-compatible rollout; distributional fidelity is assessed separately.

\subsection{Conditional Diffusion and Generative Baselines}

The DDPM corrupts standardized priors according to
\begin{equation}
q(\mathbf{z}_s\mid\mathbf{z}_0)
=
\mathcal{N}\!\left(
\sqrt{\bar{\alpha}_s}\mathbf{z}_0,
(1-\bar{\alpha}_s)\mathbf{I}
\right).
\end{equation}

A multilayer denoiser
\begin{equation}
\epsilon_\theta(\mathbf{z}_s,s,c)
\end{equation}
receives a sinusoidal diffusion-step embedding and a learned force-tier embedding and minimizes noise-prediction mean squared error.

We evaluate hidden widths 64, 128, and 256, containing 20,525, 40,749, and 105,773 parameters, respectively. Training uses 1,200 optimizer steps, batch size 128, AdamW, a cosine schedule, and 64 diffusion steps.

Baselines are a conditional VAE with 40,989 parameters and eight latent channels, a per-tier full-covariance Gaussian mixture with at most four components, and a per-tier jittered bootstrap with standardized noise $\sigma=0.03$. These non-neural references are important because the source prior is low-dimensional and smooth; successful diffusion must therefore justify itself against strong simple alternatives rather than against only weaker neural baselines.

\begin{table}[t]
\caption{Primary Experimental Matrix}
\label{tab:matrix}
\centering
\begin{tabular}{@{}lr@{}}
\toprule
Item & Value \\
\midrule
Reference train / test / blind & 512 / 96 / 96 \\
Force tiers & 5 \\
Training seeds & 5 \\
Primary generators & 6 \\
DDPM hidden widths & 64 / 128 / 256 \\
DDPM training sizes & 64 / 128 / 256 / 512 \\
Generated samples per tier/split/seed & 64 \\
Generated Basilisk trajectories & 28,800 \\
Permutation replicates per MMD test & 199 \\
Seed--tier--split metric rows & 450 \\
\bottomrule
\end{tabular}
\end{table}

\subsection{Distributional Metrics and Statistical Units}

Parameter metrics operate on standardized codec values. Trajectory metrics operate on nine standardized summaries: duration, ground range, path length, peak altitude, minimum altitude, mean speed, 95th-percentile speed, mean acceleration, and 95th-percentile acceleration. Standardization is fit independently within each force tier on training trajectories.

We report unbiased radial-basis-function MMD with pooled median bandwidth, 199-label permutation $p$ values, five-fold random-forest classifier two-sample test (C2ST) AUC, mean marginal Wasserstein distance, and nearest-training-prior distance. For each generated set and reference tier, equal sample counts are selected without replacement before MMD and C2ST.

The primary inferential unit is a seed--force-tier block ($n=25$), not an individual trajectory. Paired Wilcoxon tests \cite{wilcoxon1945} compare matched blocks. No multiplicity correction is applied to exploratory per-tier diagnostics. The prespecified scaling contrasts are 64 versus 512 trajectories on the test and blind splits.

\subsection{Downstream Diagnostic I: Sparse Observations}

The first downstream diagnostic estimates 64-step Earth-centered inertial (ECI) position and velocity sequences from noisy ECI positions. Smoothing masks retain randomly selected 10\%, 25\%, or 50\% of states and always include both endpoints. Forecast masks retain the first 25\% or 50\% of states. Independent isotropic Gaussian position noise has standard deviation 100 or 1000~m. Metrics are computed only at unobserved smoothing states or strictly after the final forecast observation.

The 512 training trajectories are partitioned by a stable trajectory-ID hash into 398 fitting and 114 calibration trajectories. The original 96 test and 96 shifted-blind trajectories remain held out.

Classical references are linear interpolation/extrapolation, a constant-velocity RTS smoother, and central-gravity Runge--Kutta propagation initialized from the RTS state. The last observed state anchors forecasts; the median observed epoch anchors smoothing propagation.

Each learned model receives noisy positions, the binary mask, normalized time, duration, declared noise level, force tier, and a complete reference trajectory. A conditional temporal U-Net and a three-layer, eight-head Transformer learn corrections to RTS. A second temporal U-Net learns corrections to the two-body trajectory. All models use width 128, 1,800 AdamW updates, batch size 64, 48 cosine diffusion steps, an $\mathbf{x}_0$ reconstruction objective, and DDIM sampling with $\eta=0.15$. Five seeds produce 16 samples per case. During reverse diffusion, observed position channels receive a soft 0.9 measurement-consistency update.

For each architecture, observation regime, and noise level, split-conformal calibration expands deviations about the ensemble mean using only the 114 calibration trajectories. Each trajectory contributes one nonconformity score: its 90th-percentile coordinate error-to-interval ratio. An outer finite-sample conformal quantile targets this criterion over exchangeable trajectories without treating correlated coordinates as independent calibration units.

We report position and velocity RMSE of the ensemble mean, energy score, 50/90/95\% coordinate coverage, 90\% radial sharpness, measurement consistency, central-gravity residual, and the fraction satisfying a local custody gate of position error below 100~km and radial 90\% spread below 200~km. This gate is a controlled decision surrogate rather than an operational custody definition.

\subsection{Downstream Diagnostic II: Ambiguous Tracking}

The second downstream diagnostic replaces Cartesian position observations with nonlinear measurements from eight rotating spherical-Earth ground stations. At an observation epoch, the station with maximum elevation emits a subset of
\begin{equation}
\mathbf{y}_k=
\left[
\rho,\dot{\rho},
\operatorname{atan2}(e,n),
\operatorname{atan2}\!\left(u,\sqrt{e^2+n^2}\right)
\right],
\end{equation}
where $(e,n,u)$ is the local station-frame displacement.

Range-only and angles-only regimes observe 10\% of the 64 epochs. A mixed regime observes range, range rate, azimuth, and elevation at 25\% density. All observations lie in the first half arc, and the terminal epoch is an unobserved custody handoff. Nominal standard deviations are 100~m, 0.2~m/s, and $0.01^\circ$; stressed values are 1000~m, 2~m/s, and $0.1^\circ$.

Each case receives two catalog hypotheses. The first is a noisy catalog state near the simulated object. For angles-only tracking, the alternate rescales the first-station line of sight; for range-only tracking, it rotates that line of sight by $2.5^\circ$. These alternatives are generated before estimation and are available identically to all methods. Likelihood weights are tempered by the number of measurements; a 10\% robust association prior prevents premature deletion in the explicitly ambiguous regimes.

This construction evaluates recovery of a declared catalog ambiguity. It does not establish that the learned ensemble discovers all modes of an unknown operational posterior.

\subsection{Classical Estimators, Physics-Residual Diffusion, and Calibration}

The classical comparators are a central-gravity EKF with RTS smoothing, a sigma-point UKF smoother, regularized nonlinear batch orbit determination, and a two-start batch candidate ensemble. Each catalog component is independently refined against the common observation packet.

The physics-residual U-Net learns a six-channel trajectory correction around each refined component. It has width 160 and is trained for 1,800 updates with 48 diffusion steps; reverse sampling emits 16 trajectories. A full-split architecture pilot also trained a three-layer, eight-head Transformer. Its test energy score was 0.01237 versus 0.01011 for the U-Net, so the prespecified 100-seed matrix retained the U-Net and reports the Transformer as a negative architecture ablation.

Residual shrinkage is selected independently per regime/noise stratum from
\begin{equation}
\{0,0.25,0.5,0.75,1\}
\end{equation}
by minimum energy score on the 114 calibration trajectories. A separate finite-sample calibration multiplier targets trajectory-level 90\% radial coverage. Neither selection uses test or blind truth.

The six regime/noise strata are averaged within each training/measurement seed, yielding 100 paired inferential units per split. We report seed-paired Wilcoxon tests, Holm family-wise correction, and seed-bootstrap 95\% confidence intervals for mean differences.

Recovery metrics include energy score, ensemble-mean RMSE, densest catalog-associated component RMSE, best-of-16 RMSE, two-component BIC support, and catalog-mode recall. The fixed custody label is whether modal terminal position error is at most 100~km. Each method ranks cases by its own posterior spread; false acceptance and success are evaluated at 50\% acceptance, together with selective-risk AUC, Brier score, and a balanced-accuracy threshold chosen only on calibration trajectories.

\subsection{GoDSAT-Compatible Monte Carlo Protocol}

The final end-to-end experiment freezes four 1,000-trajectory populations before production execution: seed replay, bootstrap, force-tier GMM samples repropagated with Basilisk, and constrained-diffusion samples from the frozen checkpoint. All files use ECI Cartesian SI states.

At campaign index $j$, all four methods receive the same GoDSAT stochastic seed. Paired method differences therefore use campaign index as the inferential pairing unit.

The common campaign contains 16 satellites, a 2-s simulation step, $20^\circ$ half-angle fields of view, a 10-tick detection refresh, a three-frame firm-track requirement, a 10-s identity latch, and a $90\times180$ ClaimMap. The predeclared failure threshold is custody fraction below 0.80.

The upstream GoDSAT logic is executed through a documented GNU Octave 9.4.0 compatibility port that preserves the satellite propagation, field-of-view, gradient-climb, slew, ownership, ClaimMap, and Q-mode state-machine logic while adding external ECI-trajectory input and JSON metric export. The experiment therefore evaluates the documented compatibility implementation and must not be interpreted as native MATLAB/Mapping Toolbox parity certification.

Mean intervals use 10,000 nonparametric bootstrap resamples; paired contrasts use common campaign indices.

\section{Results}

\subsection{Held-Out Interpolation}

\begin{figure*}[t]
    \centering
    \includegraphics[width=0.92\textwidth]{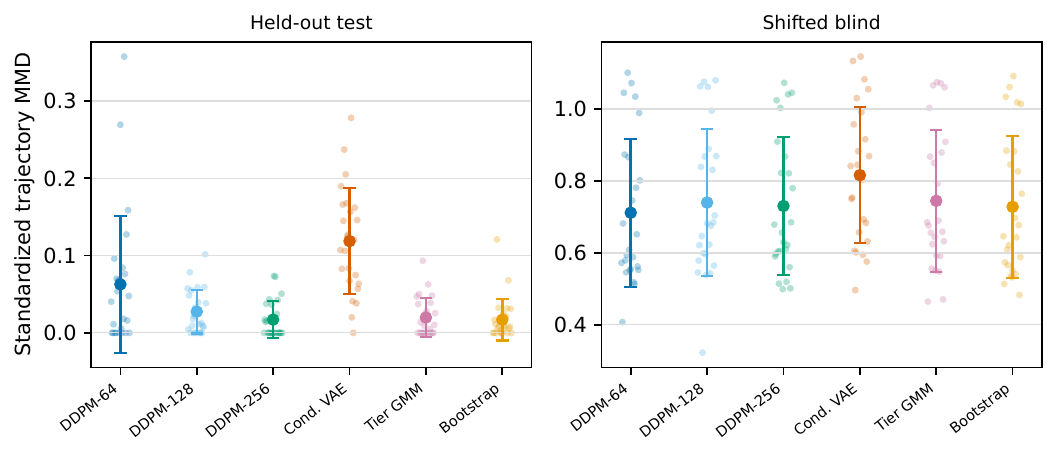}
    \caption{Trajectory MMD across 25 seed--tier blocks per model and split. Points are block values; markers and bars show mean and one standard deviation. Neural complexity does not dominate the simple baselines on held-out interpolation, and all methods degrade sharply under shift.}
    \label{fig:models}
\end{figure*}

Table~\ref{tab:primary} and Fig.~\ref{fig:models} show the primary six-model comparison. DDPM-256 reaches test trajectory MMD
\begin{equation}
0.0171\pm0.0239.
\end{equation}
It is statistically tied with jittered bootstrap
($0.0170\pm0.0268$, two-sided paired $p=0.478$)
and the per-tier GMM
($0.0198\pm0.0253$, $p=0.795$).

The GMM has the lowest parameter MMD, 0.0060, and no parameter-MMD permutation rejection in 25 blocks. The VAE is rejected in 80\% of trajectory tests and has trajectory MMD 0.1187.

\begin{table*}[t]
\caption{Primary Generator Results (Mean Over 25 Seed--Tier Blocks)}
\label{tab:primary}
\centering
\begin{tabular}{@{}lrrrrrr@{}}
\toprule
& \multicolumn{3}{c}{Held-Out Test}
& \multicolumn{3}{c}{Shifted Blind} \\
\cmidrule(lr){2-4}\cmidrule(l){5-7}
Generator
& Param. MMD
& Traj. MMD
& C2ST AUC
& Param. MMD
& Traj. MMD
& C2ST AUC \\
\midrule
DDPM-64
& 0.0320 & 0.0627 & 0.621
& 0.5071 & 0.7118 & 0.991 \\
DDPM-128
& 0.0128 & 0.0275 & 0.618
& 0.5063 & 0.7400 & 0.995 \\
DDPM-256
& 0.0122 & 0.0171 & 0.641
& 0.5125 & 0.7305 & 0.997 \\
Conditional VAE
& 0.0754 & 0.1187 & 0.757
& 0.5799 & 0.8162 & 0.998 \\
Tier GMM
& \textbf{0.0060} & 0.0198 & \textbf{0.610}
& 0.5112 & 0.7448 & 0.993 \\
Jittered bootstrap
& 0.0083 & \textbf{0.0170} & 0.620
& 0.4988 & 0.7281 & 0.994 \\
\bottomrule
\end{tabular}
\end{table*}

Adequate DDPM capacity reaches the same interpolation regime as the strong low-dimensional baselines but does not surpass them. DDPM-64 is significantly worse than the GMM on test MMD ($p=0.0355$), whereas DDPM-256 closes that gap.

The primary positive result is therefore not generator dominance. Rather, constrained diffusion can achieve competitive in-domain fidelity while producing executable trajectories and, as shown below, substantially greater distance from individual training priors than the jittered bootstrap.

\subsection{Data Scaling and Support Shift}

At fixed DDPM width 256, increasing training data monotonically reduces mean test trajectory MMD from 0.0850 with 64 trajectories, to 0.0560 with 128, 0.0223 with 256, and 0.0171 with 512. The matched 512-versus-64 change is $-0.0679$ ($p=2.69\times10^{-5}$, one-sided Wilcoxon). Parameter MMD follows the same trend, decreasing from 0.0585 to 0.0122.

\begin{figure}[t]
    \centering
    \includegraphics[width=\columnwidth]{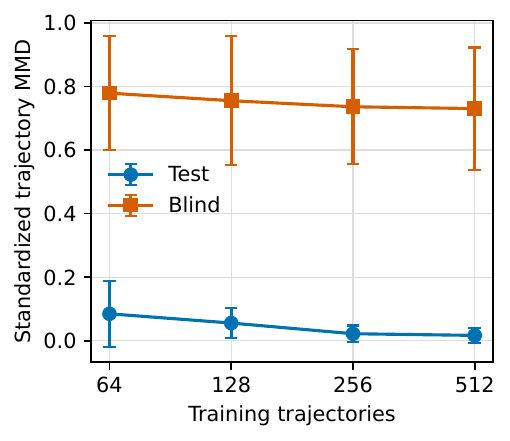}
    \caption{Sample-efficiency sweep for the 256-wide DDPM. More training data improve both regimes, but the large blind gap remains. Bars show standard deviation over 25 seed--tier blocks.}
    \label{fig:scaling}
\end{figure}

The scaling curve in Fig.~\ref{fig:scaling}, together with nearest-training-prior distance, provides evidence that the DDPM is not simply reproducing individual training rows. Median nearest-training distance for DDPM-256 is approximately 2.7 standardized units, whereas jittered bootstrap is approximately 0.10.

This distinction provides the main empirical motivation for diffusion in the unconditional augmentation setting: DDPM-256 attains held-out trajectory fidelity comparable to the strongest simple baselines while producing samples substantially farther from individual observed priors than jittered bootstrap. The increased distance should be interpreted as additional sample diversity, not as evidence of improved accuracy; bootstrap still matches DDPM-256 in test trajectory MMD.

All six primary generators fail the shifted-blind equality test in every one of 25 seed--tier blocks for both parameter and trajectory MMD. Blind parameter C2ST AUC ranges from 0.998 to 1.000, and trajectory C2ST AUC ranges from 0.991 to 0.998. DDPM-64 has the lowest mean blind trajectory MMD, 0.7118, but remains perfectly rejectable and is not a successful out-of-distribution generator.

Increasing data from 64 to 512 trajectories reduces blind trajectory MMD from 0.7791 to 0.7305 ($p=0.00139$). This detectable improvement is much smaller than the held-out improvement and does not alter the equality-test rejection. More data therefore improve interpolation substantially while failing to produce reliable extrapolation.

\begin{figure*}[t]
    \centering
    \includegraphics[width=0.94\textwidth]{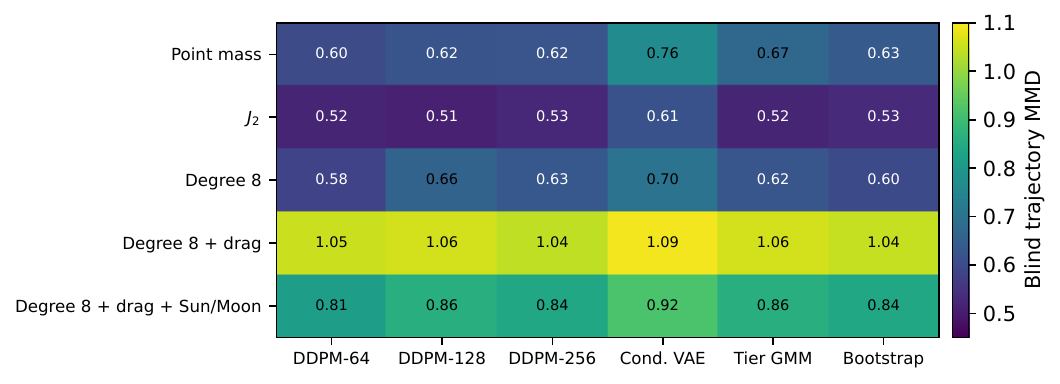}
    \caption{Shifted-blind trajectory MMD by force tier and generator. Degree-8 gravity with drag is the hardest tier for every model. No cell passes its permutation equality test.}
    \label{fig:tiers}
\end{figure*}

Figure~\ref{fig:tiers} localizes the failure. Degree-8 gravity with drag has MMD 1.02--1.09 across methods, whereas $J_2$ has MMD 0.51--0.61. The common ordering indicates that shift severity and force-tier sensitivity dominate architecture ranking.

\subsection{Constraint Representation and Lambert Consistency}

An initial codec modeled semimajor axis and eccentricity as separate bounded channels. Although each decoded marginal remained within its prescribed range, rare combinations produced perigees below the physical support and extreme drag trajectories. The final codec replaces semimajor axis with perigee altitude and derives $a$ analytically.

All 28,800 final generated samples then propagated successfully, with generated perigees between 271 and 1,398~km, within training support. This ablation demonstrates that successful simulator propagation alone is an insufficient admissibility criterion: physically meaningful constraints must also be encoded in the generative parameterization.

The Lambert benchmark uses all 800 fixed cases, with zero solver failures and no omitted trajectories. Table~\ref{tab:lambert} compares each reference trajectory and its endpoint-conditioned diffusion trajectory with the same matched Lambert arc. Exact endpoint projection succeeds to numerical precision, but the generated interior trajectories differ substantially from the corresponding two-body transfer.

\begin{table*}[t]
\caption{Matched Endpoint-Conditioned Lambert Consistency Benchmark}
\label{tab:lambert}
\centering
\begin{tabular}{@{}lrrrr@{}}
\toprule
Method
& Mean pos.
& Median pos.
& 95th pos.
& Mean vel. \\
& RMSE (km)
& RMSE (km)
& RMSE (km)
& RMSE (m/s) \\
\midrule
Reference trajectory
& 5.741 & 1.470 & 7.642 & 7.406 \\
Endpoint-conditioned diffusion
& 29.200 & 26.746 & 55.187 & 263.299 \\
\bottomrule
\end{tabular}
\end{table*}

The paired diffusion-minus-reference position-RMSE difference is $+23.459$~km with a trajectory-bootstrap 95\% confidence interval of $[+18.014,+26.836]$~km. A one-sided Wilcoxon test for greater diffusion error gives $p<10^{-127}$, and diffusion has lower position RMSE in only 5 of 800 cases.

The maximum Lambert terminal error is $2.10\times10^{-4}$~m and the maximum matched endpoint mismatch is $4.66\times10^{-10}$~m. Endpoint conditioning is therefore satisfied, but endpoint equality does not make the generated interior equivalent to the matched two-body transfer.

Because several reference tiers include non-Keplerian perturbations, this comparison should not be interpreted as an absolute ranking of physical accuracy. Its purpose is narrower: to show that satisfying boundary conditions does not by itself establish Lambert-consistent interior dynamics.

\subsection{Downstream Diagnostic: Sparse Observations}

\begin{figure*}[t]
    \centering
    \includegraphics[width=0.96\textwidth]{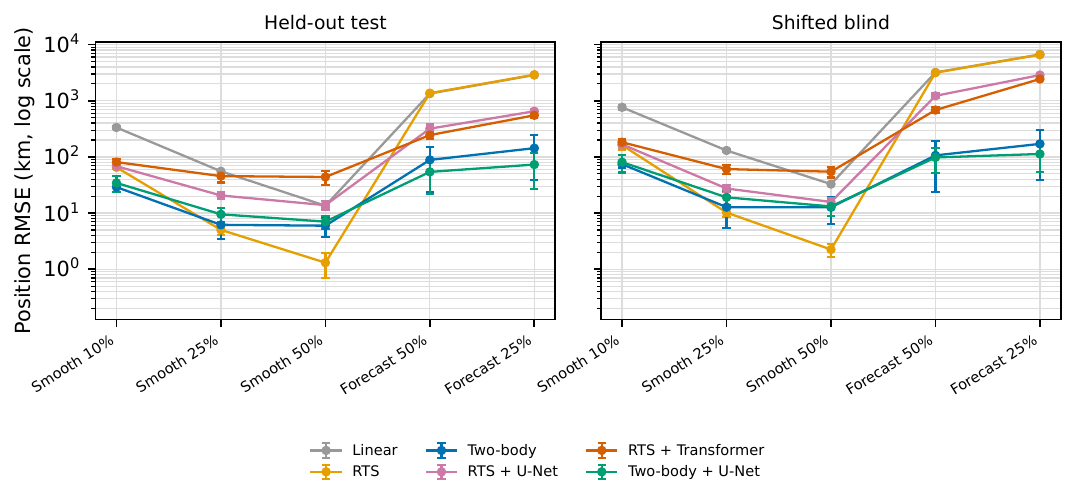}
    \caption{Position RMSE by observation regime. Points and bars show means and one standard deviation over five seeds and two noise levels. Physics-residual diffusion is strongest for forecasting, while classical estimators dominate dense smoothing.}
    \label{fig:posterior-regimes}
\end{figure*}

Table~\ref{tab:posterior} and Fig.~\ref{fig:posterior-regimes} expose a strongly regime-dependent result. On held-out forecasting, the two-body-residual U-Net reduces position RMSE from 143.2 to 73.1~km with 25\% observations and from 88.5 to 54.1~km with 50\% observations. The matched reductions are 48.9\% ($p=9.77\times10^{-4}$, 10 seed--noise pairs) and 38.8\% ($p=0.00977$).

The direct Transformer also improves substantially over RTS forecasting but remains worse than the two-body reference. In smoothing, RTS or two-body propagation remains best; learned residuals add error when observations already constrain the complete arc.

\begin{table*}[t]
\caption{Sparse-Observation Position RMSE and Calibrated Coverage (Means Over Five Seeds and Two Noise Levels)}
\label{tab:posterior}
\centering
\begin{tabular}{@{}llrrrrr@{}}
\toprule
Split
& Regime
& RTS (km)
& Two-body (km)
& Transformer (km)
& Residual U-Net (km)
& Residual 90\% cov. \\
\midrule
Test
& Forecast 25\%
& 2909.4 & 143.2 & 551.0 & \textbf{73.1} & 0.971 \\
Test
& Forecast 50\%
& 1366.1 & 88.5 & 244.8 & \textbf{54.1} & 0.967 \\
Test
& Smooth 10\%
& 65.3 & \textbf{29.0} & 80.9 & 34.4 & 0.976 \\
Test
& Smooth 25\%
& \textbf{5.0} & 6.2 & 45.5 & 9.5 & 0.965 \\
Test
& Smooth 50\%
& \textbf{1.3} & 6.0 & 43.9 & 7.0 & 0.966 \\
\midrule
Blind
& Forecast 25\%
& 6619.4 & 170.8 & 2440.7 & \textbf{112.8} & 0.973 \\
Blind
& Forecast 50\%
& 3160.8 & 106.6 & 683.9 & \textbf{97.9} & 0.945 \\
Blind
& Smooth 10\%
& 162.1 & \textbf{74.3} & 182.6 & 80.8 & 0.977 \\
Blind
& Smooth 25\%
& \textbf{10.2} & 12.7 & 60.7 & 19.1 & 0.975 \\
Blind
& Smooth 50\%
& \textbf{2.2} & 12.7 & 54.6 & 13.1 & 0.959 \\
\bottomrule
\end{tabular}
\end{table*}

Across all test conditions, the residual U-Net has 35.6~km mean RMSE versus 54.6~km for two-body propagation, 90\% coordinate coverage of 0.969, 147.7~km sharpness, and local custody fraction 0.764. On blind shift, its RMSE is 64.7~km, coverage is 0.966, and sharpness is 266.1~km. This over-coverage is the cost of trajectory-level calibration.

The corresponding blind forecast reductions are 33.9\% and 8.2\%, but neither is significant at the regime level ($p=0.0654$ and $p=0.423$). The residual model's overall paired advantage over two-body propagation is also not significant because it loses the dense-smoothing blocks.

The strongest result in this diagnostic is therefore an in-domain forecasting improvement after classical two-body propagation has removed the dominant orbital structure, together with conservative calibrated uncertainty. The experiment does not support general superiority over classical estimation.

\begin{figure*}[t]
    \centering
    \includegraphics[width=0.92\textwidth]{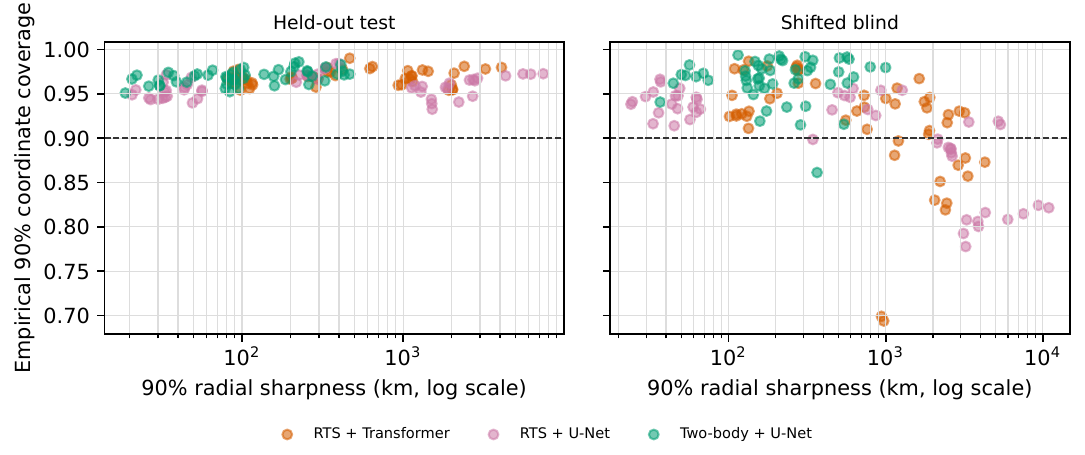}
    \caption{Empirical coverage versus interval sharpness for learned trajectory ensembles. Each point is a seed--regime--noise block; the dashed line is nominal 90\% coverage. Calibration is stable on test, while direct RTS-residual models under-cover under shift.}
    \label{fig:posterior-calibration}
\end{figure*}

\subsection{Downstream Diagnostic: Ambiguous Tracking}

\begin{figure*}[t]
    \centering
    \includegraphics[width=0.94\textwidth]{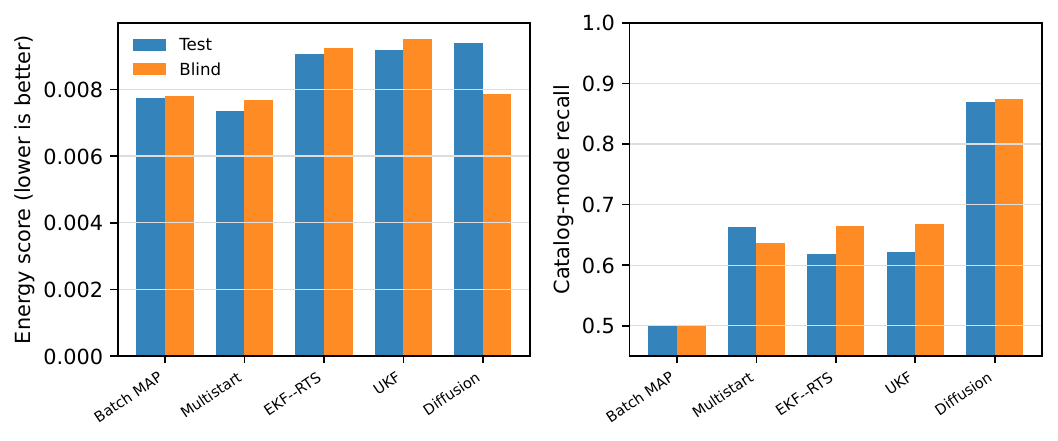}
    \caption{Proper scoring and catalog-mode recovery under nonlinear tracking. Diffusion has the highest mode recall, while multistart retains the best mean energy on both test and blind splits.}
    \label{fig:multimodal}
\end{figure*}

Table~\ref{tab:multimodal} and Fig.~\ref{fig:multimodal} separate uncertainty representation from point estimation. Physics-residual diffusion retains the declared catalog alternatives in 0.870 of test cases and 0.874 of blind cases; multistart reaches 0.664 and 0.636, respectively. Two-component BIC support is 0.629 on test and 0.604 on blind.

Against multistart, the mode-recall advantage is 0.206 on test (95\% CI $[0.203,0.209]$) and 0.238 on blind ($[0.235,0.241]$), with Holm-adjusted $p<1.1\times10^{-16}$ in both splits. Calibrated trajectory-level coverage is 0.923 and 0.914, close to the 0.90 target without calibrating on held-out trajectories.

\begin{table*}[t]
\caption{Nonlinear Multimodal Results (Means Over 100 Seeds, Three Regimes, and Two Noise Levels)}
\label{tab:multimodal}
\centering
\begin{tabular}{@{}llrrrrr@{}}
\toprule
Split
& Estimator
& Modal RMSE (km)
& Best-of-16 (km)
& Energy
& Mode recall
& Traj. 90\% cov. \\
\midrule
Test
& Batch MAP
& \textbf{13.39} & 13.39 & 0.00773 & 0.500 & 0.940 \\
Test
& Multistart
& 13.53 & \textbf{11.09} & \textbf{0.00735} & 0.664 & 0.933 \\
Test
& EKF--RTS
& 19.33 & 25.54 & 0.00907 & 0.619 & 0.918 \\
Test
& UKF smoother
& 19.52 & 26.72 & 0.00919 & 0.622 & 0.924 \\
Test
& Physics diffusion
& 19.33 & 12.60 & 0.00939 & \textbf{0.870} & 0.923 \\
\midrule
Blind
& Batch MAP
& 13.51 & 13.51 & 0.00780 & 0.500 & 0.869 \\
Blind
& Multistart
& \textbf{13.46} & 12.77 & \textbf{0.00768} & 0.636 & 0.854 \\
Blind
& EKF--RTS
& 19.37 & 28.87 & 0.00924 & 0.666 & 0.957 \\
Blind
& UKF smoother
& 19.86 & 30.25 & 0.00952 & 0.668 & 0.955 \\
Blind
& Physics diffusion
& 16.38 & \textbf{12.25} & 0.00787 & \textbf{0.874} & 0.914 \\
\bottomrule
\end{tabular}
\end{table*}

On blind seeds, diffusion improves energy over EKF by 0.00137
(95\% CI $[0.00130,0.00143]$ in absolute reduction)
and UKF by 0.00164
($[0.00158,0.00170]$),
with Holm-adjusted $p<1.1\times10^{-16}$ for both.

It is worse in mean energy than batch MAP by 0.00007
(CI $[-0.00002,0.00016]$)
and multistart by 0.00019
($[0.00011,0.00028]$).

Blind best-of-16 error improves over batch MAP by 1.264~km
(CI $[1.112,1.410]$, adjusted $p=6.1\times10^{-16}$)
and multistart by 0.523~km
($[0.385,0.661]$, $p=2.5\times10^{-8}$).

On test seeds, best-of-16 improves over batch MAP by 0.784~km
($[0.575,0.992]$, $p=1.1\times10^{-8}$),
but is 1.510~km worse than multistart; energy is worse than every reference. Point recovery therefore does not support universal superiority.

Custody provides the clearest negative result. At matched 50\% acceptance, diffusion false-accept rate is 0.0455 on test, versus 0.0152 for multistart, 0.0197 for UKF, 0.0241 for EKF, and 0.0291 for batch MAP. Its selective-risk AUC is likewise worst on test at 0.0480.

On blind data, its false-accept rate is 0.0424, statistically tied with batch MAP at 0.0428
(paired difference $-0.0004$, CI $[-0.0031,0.0022]$),
worse than multistart at 0.0398,
and worse than both Gaussian smoothers at 0.0249--0.0262.

\begin{figure}[t]
    \centering
    \includegraphics[width=\columnwidth]{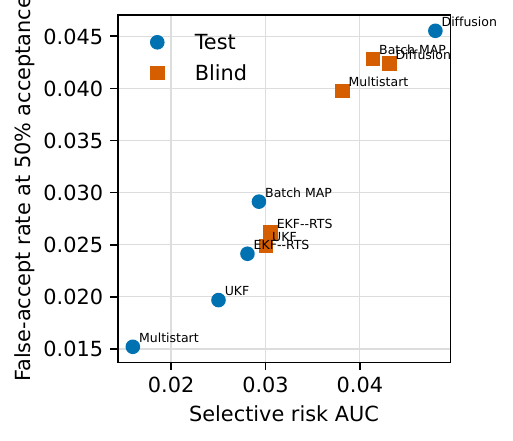}
    \caption{Custody false acceptance and selective-risk AUC. Lower is better on both axes. The diffusion ensemble does not dominate the classical references.}
    \label{fig:custody}
\end{figure}

Figure~\ref{fig:custody} therefore demonstrates that calibrated multimodal coverage does not automatically yield better operational ranking.

Diffusion therefore improves retention of the constructed catalog alternatives and selected best-of-set errors, but not point-estimation quality or custody ranking. Its strongest advantage in this experiment is representational: it can maintain multiple predeclared hypotheses that single-solution estimators tend to compress. This should not be interpreted as evidence that diffusion discovers arbitrary posterior modes that were not represented in the candidate construction.

\subsection{Fresh-Split Capacity--Budget Confirmation}

As a robustness check on the negative custody result, we test whether it can be explained by insufficient model capacity or training budget using a second independent 800-trajectory Basilisk corpus generated before evaluation.

A complete factorial crosses 100 fixed training seeds with U-Net widths 160, 320, and 640 and checkpoints at 1,800, 3,600, and 7,200 updates. These widths contain 550,886, 2,117,286, and 8,322,086 parameters, respectively. Each seed--width model is trained once to 7,200 updates and evaluated at nested checkpoints. Calibration remains confined to the training partition, and seed is the inferential unit for all nine cells on both splits.

\begin{figure*}[t]
    \centering
    \includegraphics[width=0.98\textwidth]{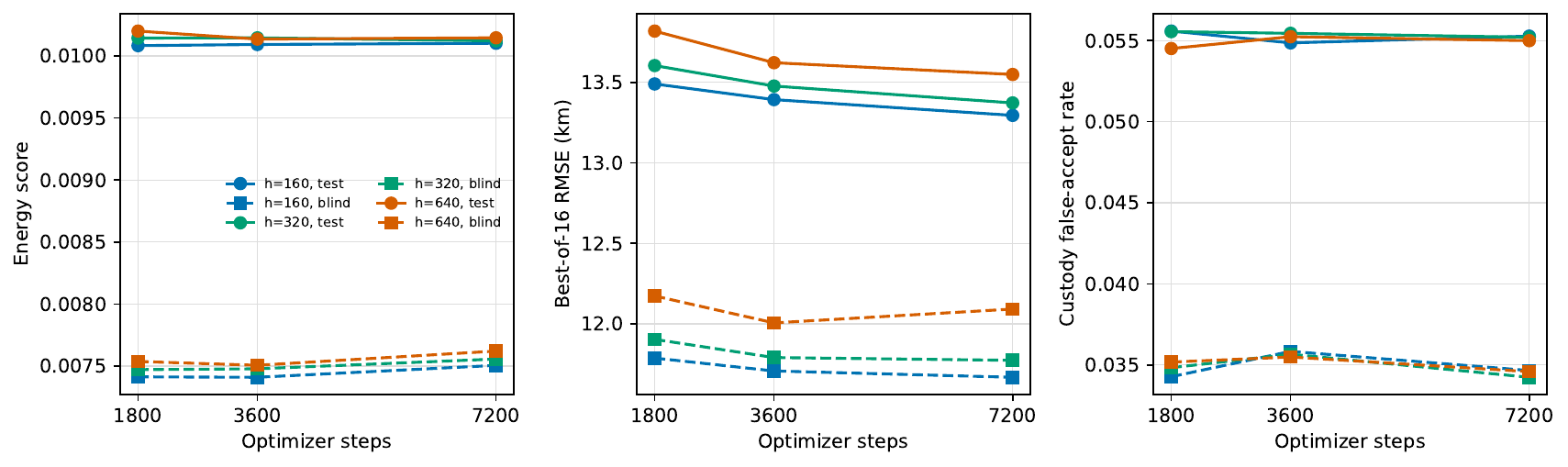}
    \caption{Fresh-split capacity--budget scaling. Solid circles denote test and dashed squares blind. Additional updates modestly reduce best-of-16 error; increased width does not improve energy or custody and generally worsens set recovery.}
    \label{fig:nonlinear-scaling}
\end{figure*}

Table~\ref{tab:capacity} and Fig.~\ref{fig:nonlinear-scaling} summarize the result. At width 160, increasing from 1,800 to 7,200 updates lowers test best-of-16 RMSE from 13.490 to 13.295~km, a paired reduction of 0.195~km
(95\% CI $[0.146,0.246]$~km; Holm $p=1.34\times10^{-8}$),
and lowers blind error from 11.787 to 11.668~km, a 0.119-km reduction
(CI $[0.062,0.176]$~km; $p=0.00368$).

At 7,200 updates, however, width 640 is worse than width 160 by 0.254~km on test and 0.424~km on blind. No direct capacity or budget contrast in false-accept rate, selective-risk AUC, balanced accuracy, or custody success survives split-wise Holm correction.

The fresh-split study therefore rejects the simplest undertraining explanation for the custody gap: additional optimization modestly improves best-of-set recovery, but neither increased capacity nor longer training materially improves decision-level performance.

\begin{table*}[t]
\caption{Fresh-Split Scaling Results (Means Over 100 Seeds)}
\label{tab:capacity}
\centering
\begin{tabular}{@{}rrrrrrrr@{}}
\toprule
Width
& Steps
& \multicolumn{3}{c}{Test}
& \multicolumn{3}{c}{Blind} \\
\cmidrule(lr){3-5}\cmidrule(lr){6-8}
&
& Energy
& Best-16 (km)
& False accept
& Energy
& Best-16 (km)
& False accept \\
\midrule
160 & 1800
& 0.010083 & 13.490 & 0.0556
& 0.007413 & 11.787 & 0.0343 \\
160 & 7200
& 0.010103 & \textbf{13.295} & 0.0553
& 0.007506 & 11.668 & 0.0347 \\
320 & 1800
& 0.010144 & 13.604 & 0.0556
& 0.007471 & 11.904 & 0.0348 \\
320 & 7200
& 0.010122 & 13.372 & 0.0552
& 0.007557 & 11.773 & 0.0342 \\
640 & 1800
& 0.010201 & 13.819 & \textbf{0.0545}
& 0.007537 & 12.173 & 0.0352 \\
640 & 7200
& 0.010147 & 13.549 & 0.0550
& 0.007621 & 12.092 & 0.0346 \\
\bottomrule
\end{tabular}
\end{table*}

Test false-accept remains 0.0545--0.0556 across all nine diffusion cells, compared with 0.0212 for multistart on the same fresh corpus. Blind diffusion false-accept is 0.0342--0.0358 versus 0.0214--0.0309 for the references. Longer training modestly improves best-of-set recovery, whereas a 15.1-fold increase in capacity does not rescue energy or custody performance.

\subsection{GoDSAT Population-Sensitivity Evaluation}

The final campaign executes 4,000 GoDSAT-compatible runs, 1,000 per frozen population, with zero crashes, retries, or missing metric values. Table~\ref{tab:godsat-final} reports the primary outcomes.

Failure fractions are high under the predeclared 0.80 custody threshold for every population and are interpreted only as conditional frequencies under this synthetic stress campaign.

\begin{table*}[t]
\caption{GoDSAT-Compatible Monte Carlo Results by Input Population}
\label{tab:godsat-final}
\centering
\begin{tabular}{@{}lrrrr@{}}
\toprule
Method
& $N$
& Mean custody [95\% CI]
& Failure fraction [95\% CI]
& Mean handoffs [95\% CI] \\
\midrule
Seed replay
& 1000
& 0.4697 [0.4513, 0.4879]
& 0.862 [0.840, 0.883]
& 2.670 [2.544, 2.797] \\
Bootstrap
& 1000
& 0.4609 [0.4422, 0.4797]
& 0.862 [0.841, 0.883]
& 2.742 [2.609, 2.877] \\
Force-tier GMM
& 1000
& \textbf{0.5177 [0.5003, 0.5356]}
& \textbf{0.848 [0.825, 0.870]}
& 2.507 [2.401, 2.612] \\
Constrained diffusion
& 1000
& 0.4631 [0.4444, 0.4817]
& 0.861 [0.840, 0.882]
& 2.690 [2.562, 2.821] \\
\bottomrule
\end{tabular}
\end{table*}

Paired custody differences for constrained diffusion are
$-0.00666$ versus seed replay
(95\% CI $[-0.03190,0.01867]$),
$+0.00215$ versus bootstrap
($[-0.02346,0.02819]$),
and $-0.05464$ versus the force-tier GMM
($[-0.08052,-0.02860]$).

Diffusion is therefore statistically indistinguishable from seed replay and bootstrap on mean custody, whereas the GMM population produces custody higher by approximately 5.46 percentage points.

The force-tier GMM therefore changes estimated mean custody by approximately 5.46 percentage points relative to constrained diffusion under the same campaign configuration. This difference must not be interpreted as evidence that the GMM is a better generator: each method induces a different synthetic evaluation population. Rather, the result demonstrates that trajectory-population construction is itself a material modeling choice in data-scarce constellation Monte Carlo analysis. The thresholded failure contrasts all include zero, making continuous custody more discriminating in this campaign.

Secondary metrics support the same caution. Diffusion has mean handoff latency 80.438~s, mean Q-mode reacquisition 83.031~s, and mean ClaimMap conflicts 0.016. Relative to the GMM, its unconditional handoff latency is $+9.581$~s
(95\% CI $[1.279,17.718]$);
relative to bootstrap, its conflict difference is $+0.011$
($[0.002,0.020]$).

Because campaigns with no completed handoff or reacquisition contribute zero under the frozen convention, these unconditional means are interpreted together with event incidence. The prespecified stopping rule triggered at 1,000 campaigns per method after custody confidence intervals narrowed sufficiently to resolve the qualitative comparison, so the optional 5,000-per-method extension was not executed.

\section{Discussion}

\subsection{What the Results Establish}

The experiments delineate a narrow but useful role for constrained diffusion. Within observed support, DDPM-256 reaches the same trajectory-fidelity regime as strong low-dimensional baselines while producing samples farther from individual training priors. Under prescribed support shift, however, every evaluated generator is readily distinguishable from the target distribution. Classical comparisons further show that generative flexibility does not remove the need for physical references: endpoint conditioning does not establish Lambert-consistent interiors, and learned residuals improve only selected estimation regimes.

Under sparse observations, diffusion is most useful after two-body propagation has already captured the dominant dynamics, while classical estimators remain preferable for dense smoothing. Under constructed catalog ambiguity, diffusion preserves predeclared alternatives better than the classical references, yet multistart retains stronger proper-score and custody behavior. The independent fresh-split scaling study makes insufficient capacity or training budget an unlikely explanation for that decision-level gap.

These results distinguish three properties that are easy to conflate:
\begin{enumerate}
    \item physical executability,
    \item distributional or uncertainty-representation quality, and
    \item downstream decision quality.
\end{enumerate}
Success on one does not imply success on the others.

The GoDSAT-compatible campaign provides the paper's central constellation-level implication: Monte Carlo conclusions can depend materially on the assumed trajectory-population model. Constrained diffusion, seed replay, and bootstrap produce similar mean custody, whereas the force-tier GMM shifts the estimate upward by approximately 5.46 percentage points under the same campaign configuration.

Because the methods generate different synthetic evaluation populations, this difference is not a ranking of generator quality. It is evidence that trajectory-population construction is itself a source of uncertainty in data-scarce constellation evaluation.

\subsection{Why Diffusion When Simple Baselines Are Strong?}

The unconditional prior space is only 13-dimensional, and the experiments show that simple statistical models are consequently strong competitors. The GMM achieves the lowest parameter MMD, and jittered bootstrap matches the strongest DDPM in held-out trajectory MMD. Diffusion is therefore not justified here by superior density estimation alone.

Its empirical value is instead twofold. First, DDPM-256 achieves held-out trajectory fidelity comparable to bootstrap and GMM sampling while producing samples substantially farther from individual training priors than jittered bootstrap. Second, the same conditional generative machinery extends to residual forecasting and multimodal candidate representation, where the task is no longer merely low-dimensional density estimation.

These results support diffusion as a flexible augmentation mechanism, not as the universally preferred generator for data-scarce orbital populations.

\subsection{Constellation-Level Use}

The intended role of constrained diffusion is to expand the conditions under which GoDSAT's existing tasking and custody logic is evaluated, not to replace its orbit propagator.

Within the declared training support, generated populations can reduce repeated use of identical seed cases and can be used to construct controlled stress-test populations spanning varied handoff geometry, observation sparsity, and catalog ambiguity. Their frequency within such synthetic populations must not be interpreted as an operational occurrence probability without external calibration.

This distinction matters because synthetic-data quality and downstream task difficulty are not the same objective. A generator that produces higher custody may simply produce an easier population. Conversely, a generator that preserves more uncertainty modes may still rank custody risk worse.

For Monte Carlo analysis, the relevant question is therefore not which generator produces the most favorable downstream metric, but whether the generated population is externally justified for the intended operating regime.

\subsection{Scope of Physical Validity}

Every unconditional generated sample is propagated by the same Basilisk force tier used to create the reference benchmark. This removes state-level inconsistency and provides exact force-tier labeling, but it establishes only consistency with that simulator.

The matched Lambert study adds a classical two-body consistency reference for endpoint-conditioned cases and shows that endpoint equality alone is insufficient evidence of interior two-body consistency. It does not rank absolute physical fidelity across force tiers, independently validate Basilisk, establish agreement with an operational population, or demonstrate extrapolation. The shifted-blind results directly reject the last claim.

The framework is therefore a learned Monte Carlo prior coupled to a simulator, not a neural replacement for orbital dynamics.

Stronger physical validation would require cross-propagation in a separately implemented high-fidelity engine and comparison with external ephemerides or controlled observations. Those evaluations are outside the computational scope of the present study and are not implied by the results reported here.

\subsection{Limitations and Threats to Validity}

All trajectories are synthetic and sampled from known ranges. The 800-trajectory corpus supports controlled comparisons but is not representative of an operational catalog, and the 19--20 objects per force tier limit per-tier resolution. Repeated training and measurement seeds reuse the same held-out objects. The reported intervals therefore measure conditional algorithmic, measurement, and campaign variation rather than uncertainty over an operational population. Accordingly, the large number of repeated seeds increases precision for the controlled benchmark but does not increase the number of independently sampled physical objects.

The observation models also simplify field conditions. They assume known noise, synchronized clocks, exact station locations, no maneuvers, no missed detections, and no systematic measurement bias. Catalog ambiguity is constructed rather than estimated from an empirical confusion distribution, and 16 diffusion samples provide limited resolution of low-probability modes. Trajectory-level conformal calibration does not guarantee simultaneous coverage of every state coordinate under covariate shift.

Basilisk serves as both the reference-data generator and unconditional rollout engine, so physical executability is not independent physical validation. The Lambert experiment evaluates one sampled diffusion trajectory per endpoint case and uses a favorable reference-informed branch rule; it is a controlled two-body consistency benchmark rather than an operational transfer-selection procedure.

The GoDSAT-compatible campaign uses a documented GNU Octave compatibility port of the upstream source rather than the native MATLAB/Mapping Toolbox runtime. The reported results therefore characterize the compatibility implementation under the frozen campaign configuration and do not constitute an empirical certification of exact native-runtime parity.

All four GoDSAT input populations derive from the same finite synthetic corpus or models fit to it. Pairing common campaign seeds controls shared simulation randomness but does not make trajectories from different generated populations physically matched.

The 1,000-per-method stopping point resolves the principal custody contrast but is not designed to estimate rare ClaimMap conflicts precisely. Neither the local custody surrogate nor the end-to-end synthetic campaign constitutes mission or operational certification.

\subsection{Follow-On Work}

Three extensions follow directly from the results.

First, broader or hierarchical training priors should be evaluated with explicit conditional coverage targets rather than with an expectation of implicit extrapolation.

Second, flow matching, normalizing flows, copulas, and autoregressive models should be compared within the same constrained latent space. The present low dimensionality may favor explicit-density approaches, and the strong GMM and bootstrap results make this comparison especially relevant.

Third, campaign-level conclusions should ultimately be calibrated against an independently defined target population, real tracking observables, separately implemented dynamics, and native MATLAB/Mapping Toolbox parity for GoDSAT where applicable.

\subsection{Reproducibility}

Code, trajectory splits, generated parameters, estimator outputs, scaling results, Lambert case metrics, GoDSAT-compatible campaign metrics, paired contrasts, frozen campaign configuration, and figure-generation scripts are available at
\url{https://github.com/oramada/godSAT_Diffusion}.

The release includes manifests, checkpoint and file hashes, seed-level records, 4,000 raw GoDSAT JSON outputs, and the machine-readable execution audit needed to reconstruct the reported comparisons.

These materials are intended to support exact computational reproduction of the reported controlled experiments. They do not provide independent scientific validation, which would require external observations, separately implemented dynamics, and native-runtime parity where applicable.

\section{Conclusion}

Constrained diffusion can generate executable orbital ensembles whose held-out trajectory statistics match strong low-dimensional baselines within observed support. In this benchmark it does not provide superior density estimation: the largest model ties a jittered bootstrap and a per-tier Gaussian mixture on held-out trajectory maximum mean discrepancy while producing samples farther from individual training priors. That additional distance is diversity, not accuracy, and it is useful only inside the declared support.

The same generators all fail a prescribed support shift. Endpoint-conditioned generation satisfies Lambert boundary conditions and does not recover the interior transfer. Physics-residual diffusion improves selected sparse forecasts and retains constructed catalog modes; it does not outperform classical estimators on dense smoothing or custody ranking. Classical astrodynamics remains the appropriate physical reference.

The constellation-level result is the population comparison, not a ranking of samplers. Across 4,000 GoDSAT-compatible campaigns, constrained diffusion is indistinguishable from seed replay and bootstrap on mean custody, whereas a force-tier Gaussian mixture shifts the estimate by approximately 5.5 percentage points under the same tasking configuration. Because the methods induce different synthetic evaluation populations, a higher custody fraction may simply indicate an easier sampled mix.

Constrained diffusion is therefore a complementary mechanism for locally supported Monte Carlo augmentation. It can reduce repeated use of identical seed cases inside declared training bounds. It is not a neural orbit propagator, not an operational posterior, and not a substitute for independently validated astrodynamics. Reliable use requires an externally justified target population and explicit handling of support shift. Trajectory-population construction is itself a first-order uncertainty in data-scarce constellation analysis and should be reported with the same care as the tasking configuration.

%%%%%%%%%%%%%%%%%%%%%%%%%%%%%%%%%%%%%%
\section*{Acknowledgments}
%%%%%%%%%%%%%%%%%%%%%%%%%%%%%%%%%%%%%%

The authors gratefully acknowledge the Computer and Information
Technology (CIT) program within the Johns Hopkins University Whiting
School of Engineering, Engineering for Professionals, which encompasses
Artificial Intelligence, Computer Science, Cybersecurity, Data
Analytics, Data Science, and Information Systems Engineering. The
support provided through this program was instrumental to this work.

In accordance with IEEE guidance, the authors disclose that Claude (Anthropic) assisted with prose drafting and editing in the Abstract, Introduction, Methodology, Results, Discussion, and Conclusion, and with implementing Python analysis code. OpenAI Codex assisted with LaTeX formatting, bibliography conversion and metadata checking, and editing of Related Work, biographies, and this acknowledgment. AI assistance also included reviewing Python analysis code for possible defects and checking mathematical notation, equations, and derivations in the Methodology and Results for internal consistency. These checks were aids to author review, not independent validation of the code or mathematics. The authors remain responsible for the scientific claims, calculations, numerical results, sources, and final content.

\bibliographystyle{IEEEtran}
\bibliography{bibliographies}

% The written 2027 instructions specify 1.25 by 1.5 inch portraits. Override
% only the biography environment; the downloaded conference class stays intact.
% Optional trim values crop each source photo to a 5:6 portrait without stretching.
\renewenvironment{biographywithpic}[3][0 0 0 0]{%
\normalsize\unitlength 1mm\bigskip\parskip=0pt\par
\rule{0pt}{45mm}\vspace{-45mm}\par
\noindent\setbox0\hbox{\includegraphics[trim=#1,clip,width=1.25in,height=1.5in]{#3}}%
\ht0=44mm\count10=\ht0\divide\count10 by\baselineskip
\global\hangindent35.75mm\global\hangafter-\count10
\hskip-35.25mm\setbox0\hbox to 35.25mm {\raise-36.6mm\box0\hss}%
\dp0=0mm\ht0=0mm\box0\noindent\textit{\textbf{#2}}}{%
\par\rmfamily\normalsize}

\newpage
\thebiography

\begin{biographywithpic}[0 56bp 0 0]{Omar Ramadan}{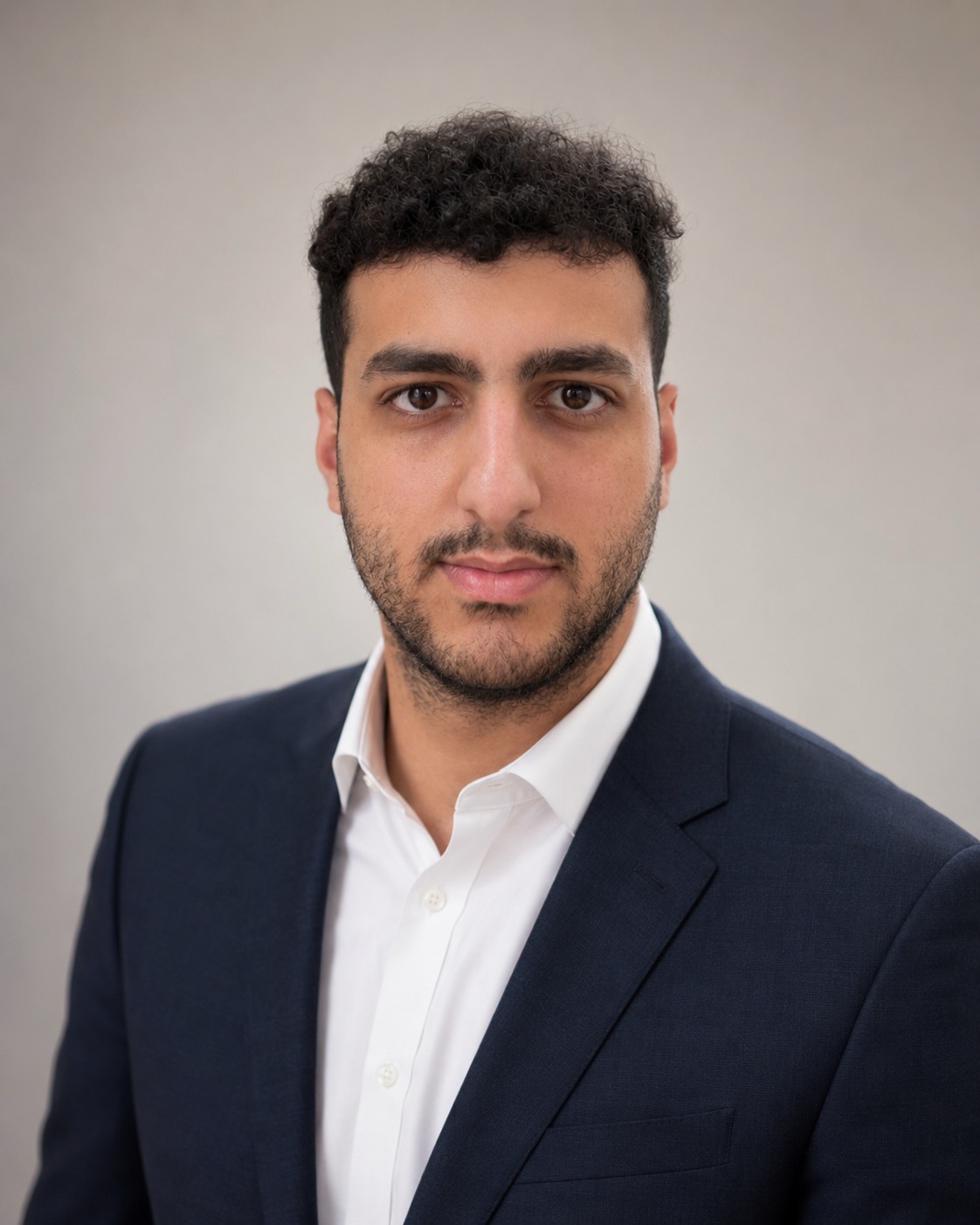}\itshape\ 
is a Data Science master's student at Johns Hopkins University researching physically grounded and scalable AI for complex environments. He has developed physics-informed neural networks and Bayesian neural networks that incorporate vector calculus and partial differential equations into trajectory models, alongside Monte Carlo methods for uncertainty quantification. He has built GPU-accelerated pipelines and custom CUDA kernels for physics-based learning. His work with graph neural networks and spectral graph theory explores how to reconstruct evidence and improve the reliability of large language model outputs. He has conducted machine-learning research at Johns Hopkins University and the University of Baltimore and published at IEEE COMPSAC 2025 and SPIE 2026. His interests span computer vision, robotics, embodied intelligence, and autonomous systems.
\end{biographywithpic}

\begin{biographywithpic}{Sam Siavoshian}{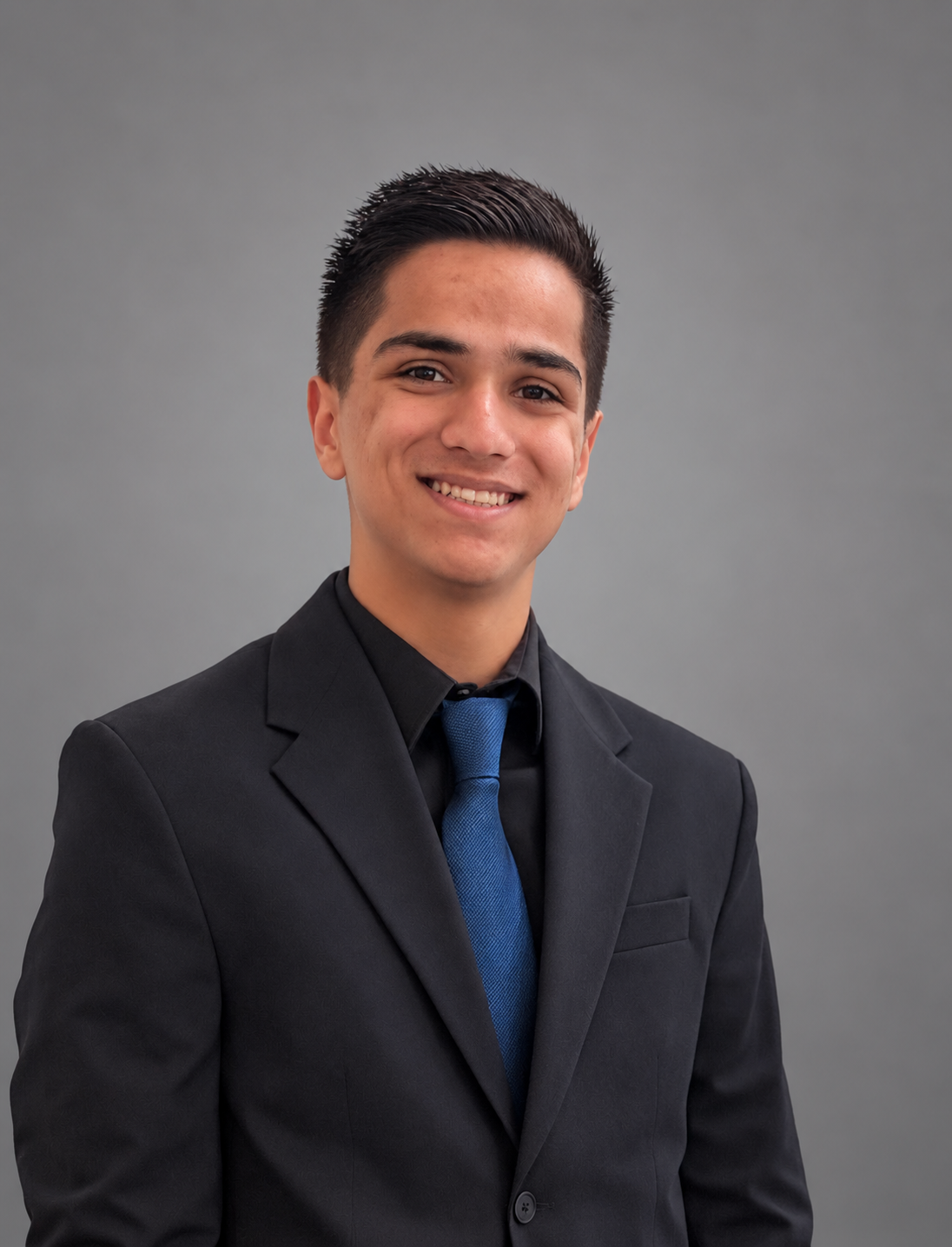}\itshape\ 
is an independent researcher, software engineer, and founder focused on artificial intelligence, machine learning, and intelligent systems. For this work, he contributed to the conceptualization of the study, the architecture of the dataset and experimental pipeline, and the training of the diffusion models. His research explores persistent memory, temporal awareness, continuous computation, and real-time AI, including work on Chronometric Injection and Continuous Stateful Intelligence. He has also built AI agents, browser-automation systems, and machine-learning products, and has contributed as a software engineer at a Y Combinator-backed startup. His broader research interests center on AI systems that maintain state, reason continuously over time, and interact with dynamic real-world environments.
\end{biographywithpic}

\begin{biographywithpic}[0 108bp 0 0]{Amir K. Saeed}{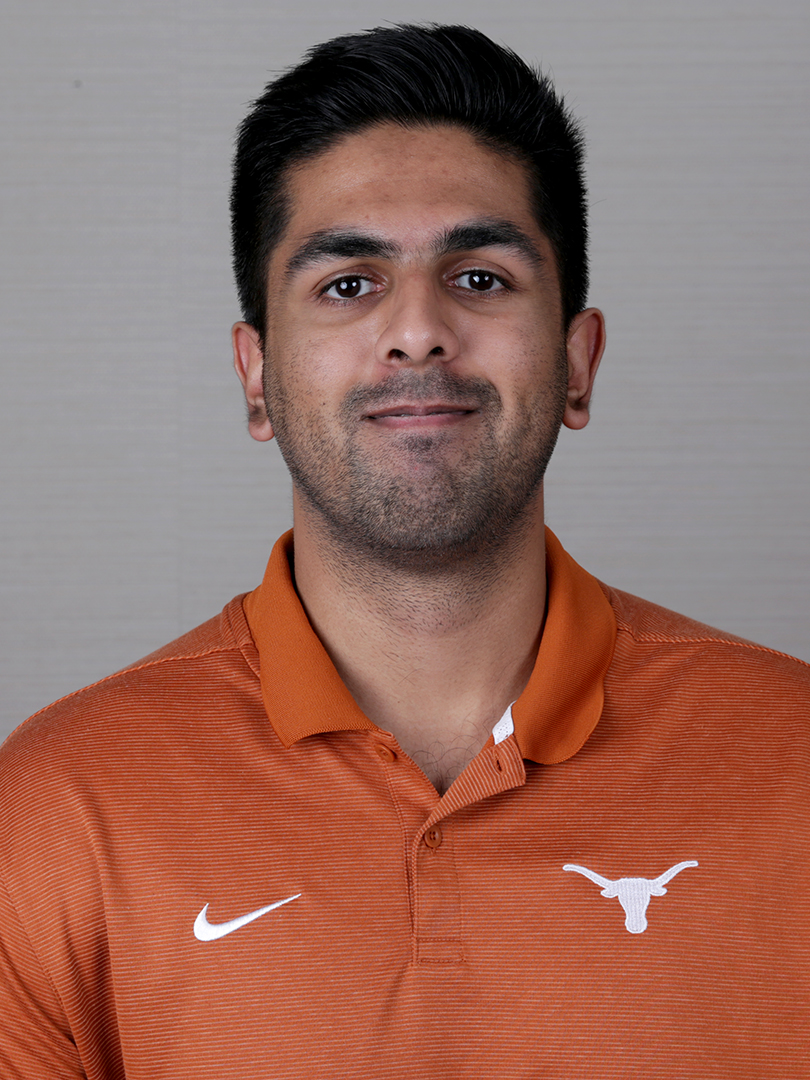}\itshape\ 
is an AI Forward Deployed Engineer at Arize AI and adjunct faculty in data science and artificial intelligence at Johns Hopkins University. He previously served as a Lead Machine Learning Scientist at Baker Hughes, contributing to energy technology projects in additive manufacturing, turbomachinery, asset management, forecasting, modeling, and emissions control. His career began in mechanical engineering, with an undergraduate degree from The University of Texas at Austin followed by a master's degree in data science from Johns Hopkins University. His expertise spans design and process engineering, data analysis, model building, and software development. His research interests include multimodal data fusion, reinforcement learning, and effective data visualization. He teaches Algorithms and Data Patterns and Representations in the Johns Hopkins Engineering for Professionals program.
\end{biographywithpic}

\begin{biographywithpic}[0 23bp 0 0]{Benjamin A. Johnson}{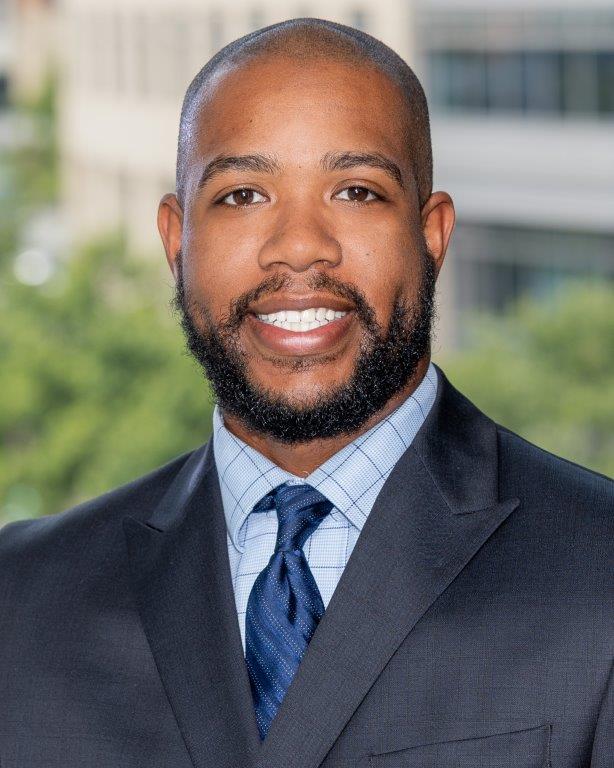}\itshape\ 
is a senior analytics manager and program manager in the Intelligence Portfolio at Accenture Federal Services. His recent research concerns data fusion and reinforcement learning for satellite tasking. He previously served for 10 years as a U.S. Army Signal Officer. He holds a master's degree in data science from Johns Hopkins University and a bachelor's degree in mathematics from the United States Military Academy at West Point.
\end{biographywithpic}

\newpage
\begin{biographywithpic}[104bp 0 104bp 0]{Amin M. E.-A. Diab}{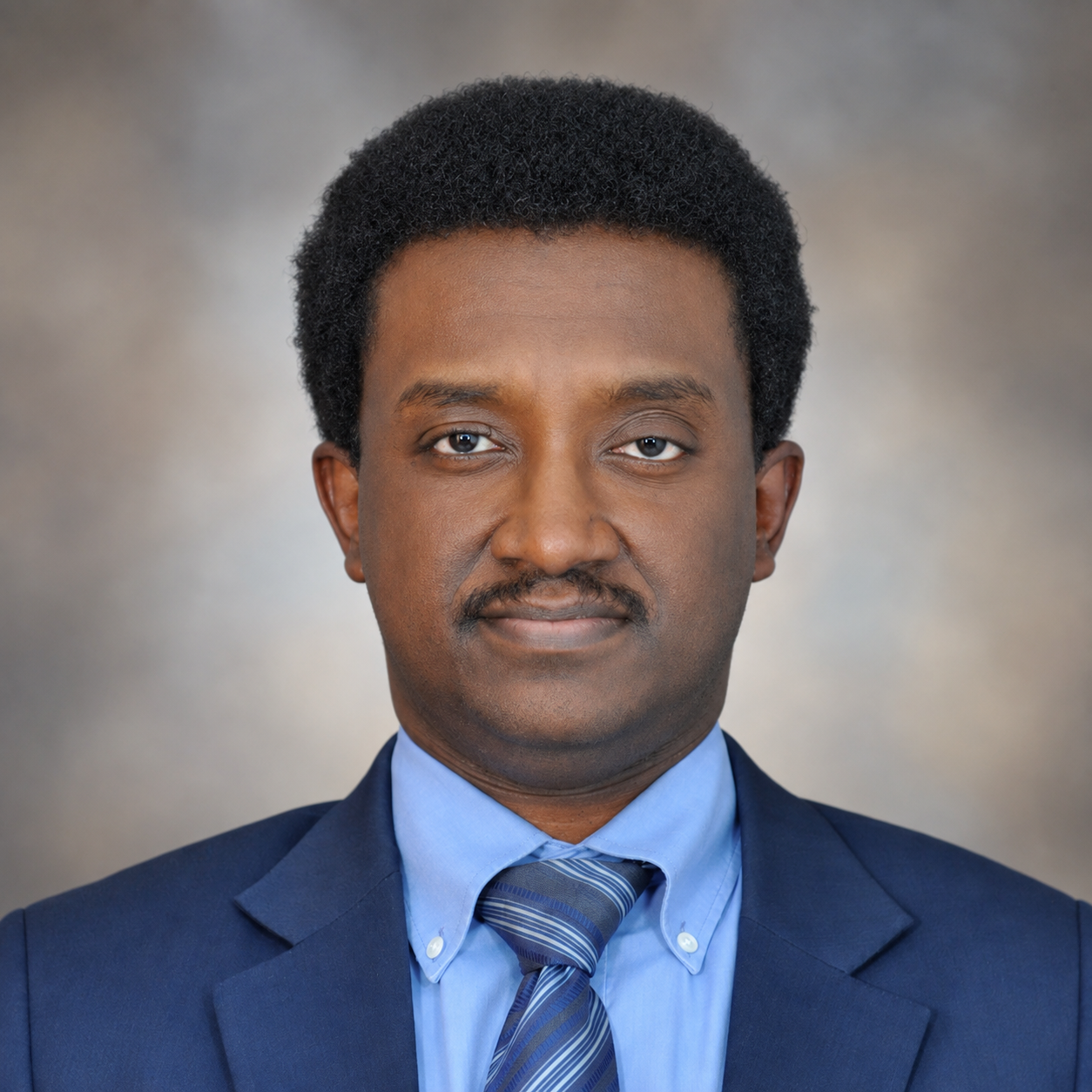}\itshape\ 
conducts artificial intelligence research at Johns Hopkins University. His work spans generative and agentic AI, reinforcement learning, transformer and multimodal models, computer vision, and scalable AI systems. He has built healthcare language models that use retrieval to augment generation, as well as reinforcement learning models and cloud-based AI pipelines for healthcare, cybersecurity, public safety, and financial analysis. His earlier research addressed photonic data communication systems, on which he has authored several articles. He holds a Ph.D. in photonics engineering, a Master of Engineering from the University of Cambridge, and a Master of Science in artificial intelligence from Johns Hopkins University.
\end{biographywithpic}

\begin{biographywithpic}[0 180bp 0 0]{Benjamin M. Rodriguez}{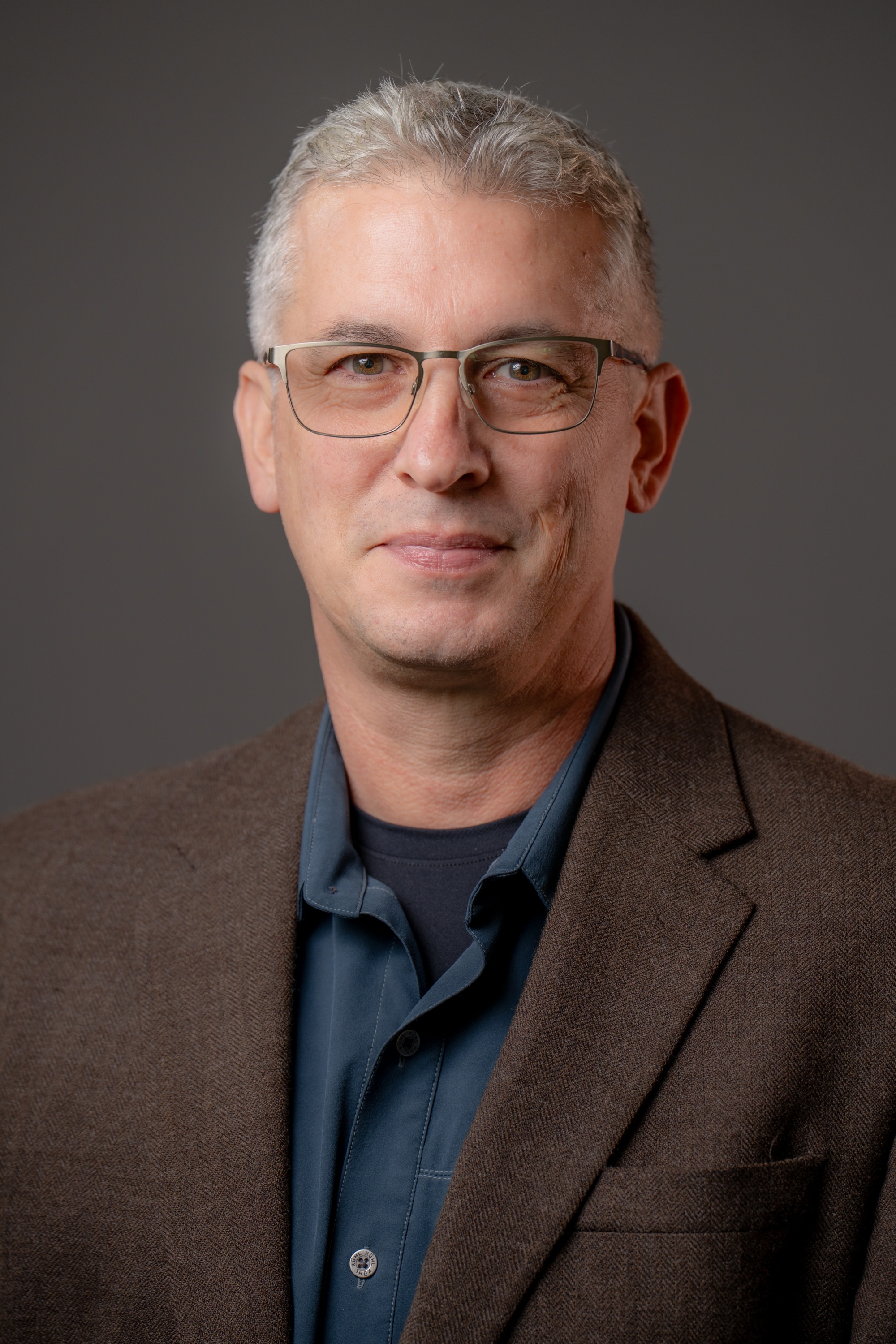}\itshape\ 
chairs the Data Analytics Engineering Program and co-chairs the Data Science Program at the Johns Hopkins Whiting School of Engineering. He is also Principal Professional Staff at the Johns Hopkins University Applied Physics Laboratory, where he leads multidisciplinary work on space-based sensing, air systems, force design modeling, and test and evaluation. His research combines data fusion, machine learning, pattern recognition, autonomous decision-making, and modeling and simulation for complex systems and national security. Through his research and teaching, he advances practical AI methods and graduate education in data science and analytics.
\end{biographywithpic}

\end{document}